%% file: aaai21.tex
\def\year{2021}\relax
\documentclass[letterpaper]{article} 
\usepackage{aaai21}  
\usepackage{times}  
\usepackage{helvet} 
\usepackage{courier}  
\usepackage[hyphens]{url}  
\usepackage{graphicx} 
\usepackage{natbib}  
\usepackage{caption} 
\usepackage{booktabs} 

\usepackage[english]{babel}
\usepackage{moresize}
\usepackage{amsmath}

\usepackage{algorithmic} 
\usepackage{balance}
\usepackage{comment}
\usepackage{paralist}
\usepackage{multirow}

\usepackage{filecontents}

\usepackage[english]{babel}
\usepackage[latin1]{inputenc}
\usepackage{mathrsfs}
\usepackage{graphicx}
\usepackage{amssymb}
\usepackage{url}
\usepackage{subfigure}
\usepackage{amsmath}
\usepackage{enumitem}
\usepackage[linesnumbered,algoruled,boxed,lined]{algorithm2e}
\usepackage{adjustbox}
\usepackage{amssymb}
\usepackage{times}
\usepackage{hyperref}

\usepackage{pgfplots}
\usetikzlibrary{pgfplots.dateplot}

\usepackage{filecontents}
\definecolor{tblue}{RGB}{31,119,180}
\definecolor{torange}{RGB}{255,127,14}
\definecolor{tgreen}{RGB}{44,160,44}
\definecolor{tred}{RGB}{214,39,40}
\definecolor{tpurple}{RGB}{148,103,189}

\usepackage{filecontents}

\newcommand{\hide}[1]{} 

\newcommand{\ie}{\textit{i}.\textit{e}.}
\newcommand{\eg}{\textit{e}.\textit{g}.}

\title{Graph-Enhanced Multi-Task Learning of Multi-Level Transition Dynamics for Session-based Recommendation}

\author{
Chao Huang$^1$, Jiahui Chen$^2$, Lianghao Xia$^2$, Yong Xu$^{2,3,4}$\thanks{Corresponding author: Yong Xu}, Peng Dai$^1$, \\\Large{\bf Yanqing Chen$^1$, Liefeng Bo$^1$, Jiashu Zhao$^5$, Jimmy Xiangji Huang$^6$}\\
}
\affiliations{
    \textsuperscript{\rm 1}JD Finance America Corporation, USA\\
    \textsuperscript{\rm 2}South China University of Technology, China, \textsuperscript{\rm 3}Peng Cheng Laboratory, China\\
    \textsuperscript{\rm 4}Communication and Computer Network Laboratory of Guangdong, China\\
    \textsuperscript{\rm 5}Wilfrid Laurier University, Canada, \textsuperscript{\rm 6}York University, Canada\\
chaohuang75@gmail.com, \{201721041314, cslianghao.xia\}@mail.scut.edu.cn, yxu@scut.edu.cn, \\ \{peng.dai, yanqing.chen, liefeng.bo\}@jd.com, jzhao@wlu.ca, jhuang@yorku.ca

}

\def\model{MTD}

\begin{document}


\maketitle

\begin{abstract}
Session-based recommendation plays a central role in a wide spectrum of online applications, ranging from e-commerce to online advertising services. However, the majority of existing session-based recommendation techniques (\eg, attention-based recurrent network or graph neural network) are not well-designed for capturing the complex transition dynamics exhibited with temporally-ordered and multi-level inter-dependent relation structures. These methods largely overlook the relation hierarchy of item transitional patterns. In this paper, we propose a multi-task learning framework with \underline{M}ulti-level \underline{T}ransition \underline{D}ynamics (\model), which enables the jointly learning of intra- and inter-session item transition dynamics in automatic and hierarchical manner. Towards this end, we first develop a position-aware attention mechanism to learn item transitional regularities within individual session. Then, a graph-structured hierarchical relation encoder is proposed to explicitly capture the cross-session item transitions in the form of high-order connectivities by performing embedding propagation with the global graph context. The learning process of intra- and inter-session transition dynamics are integrated, to preserve the underlying low- and high-level item relationships in a common latent space. Extensive experiments on three real-world datasets demonstrate the superiority of \model\ as compared to state-of-the-art baselines. 
\end{abstract}


\input{intro}
\input{solution}
\input{eval}
\input{relate}
\input{conclusion}

\section*{Acknowledgments}
We thank the anonymous reviewers for their constructive feedback and comments. This work is supported by National Nature Science Foundation of China (62072188, 61672241), Natural Science Foundation of Guangdong Province (2016A030308013), Science and Technology Program of Guangdong Province (2019A050510010). This work is also partially supported by the Natural Sciences and Engineering Research Council of Canada (NSERC) and the York Research Chairs (YRC) program.


\bibliography{refs}

\end{document}

%% file: intro.tex
\section{Introduction}
\label{sec:intro}

Personalized recommendation has attracted a lot of attention in real-life applications, to alleviate information overload on the web~\cite{xia2020multiplex}. In various recommendation scenarios, session-based recommendation has become an important component in many online services (\eg, retailing and advertising platforms)~\cite{huang2004dynamic}, to address the unavailability issue of user information in realistic scenarios (such as non-logged in customers or users without historical interactions)~\cite{quadrana2017personalizing,ren2019repeatnet,yuan2020future}. At its core is to predict the next interactive item based on a group of anonymous temporally-ordered behavior sequences of users (\eg, clicked, browsed or purchased item sequences)~\cite{liu2018stamp,wang2020global,wang2019collaborative}. To facilitate the study of session-based recommendation, many efforts have been devoted to developing various deep neural network models, by exploring correlations between the future interested item and past interacted ones, which contributes to smarter recommendations.

\begin{figure*}[t]
    \includegraphics[width=0.99\textwidth]{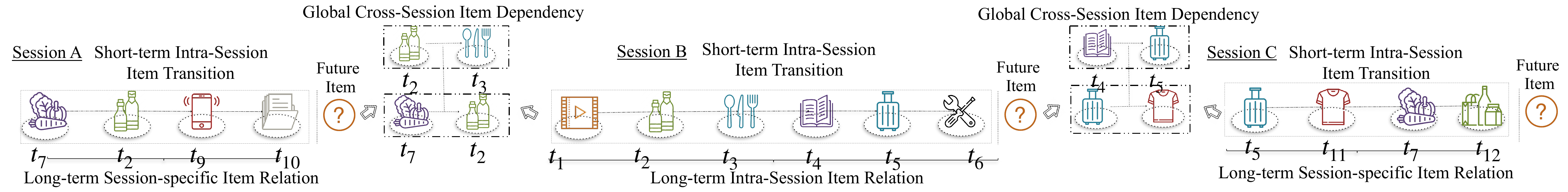}
    \vspace{-0.10in}
    \caption{Illustrated example of session-based recommendation with multi-level transition dynamics.}
    \label{fig:intro_example}
    \vspace{-0.15in}
\end{figure*}

Existing session-based recommendation methods for understanding the item transitional regularities can be grouped into several key paradigms. For example, one key research line aims to capture transitional patterns of interacted item sequence with recurrent neural network~\cite{hidasi2015session,hidasi2018recurrent}. Along this line, to aggregate sequential embeddings into a more summarized session-level representation, researchers recently propose to augment recurrent session-based recommendation frameworks with attention mechanism~\cite{li2017neural}, or rely on the memory network~\cite{liu2018stamp,wang2019collaborative}. Furthermore, another recommendation paradigm utilizes graph neural network as the item transitional relation encoder, to model long-term item dependencies within the session based on the structured relation graph~\cite{wu2019session}.

Despite their effectiveness, we argue that these methods are not sufficient to yield satisfactory recommendation results, due to their failure in encoding complex item transition dynamics which are exhibited with multi-levels in nature~\cite{song2019hierarchical}. Particularly, in the practical session-based recommendation scenarios, there exist session-specific short-term and long-term item transitions, as well as the long-range cross-session item dependencies in global context~\cite{al2018dynamic}. These different inter-correlations among items constitute the underlying multi-level item transition dynamics. As illustrated in Figure~\ref{fig:intro_example}, while item $t_7$ and $t_3$ are not directly connected within the same session, there exist implicit inter-dependency among them, due to the item transitional relationship of $t_2 \rightarrow t_3$ and $t_7 \rightarrow t_2$ in session $B$ and $A$, respectively. In such cases, items from different sessions are no longer independent. The dependent signals between interactive items may come from not only the intra-session transition regularities, but also inter-session item relations. However, to simplify the model design, most of current session-based recommender systems only explore local contextual features, while the global item transitional patterns across exogenous sessions are neglected. This restricts the capabilities of current models in capturing the hierarchical transition signals for making recommendations.

While intuitively useful to perform the joint learning of item relation structures with multi-level transition dynamics, it is non-trivial to do it well. In particular, the item dependencies across different sessions can be complex. It is not necessary that a future interactive item is more relevant to items from a recent session than one that is further away~\cite{kang2018self}. Hence, when tackling the cross-session item dependencies at various neighbor distances, the high-order relation structures exhibited with item transition patterns from a global perspective over all sessions, is necessary to be investigated in the relation embedding function. Additionally, intra-session item transition patterns vary by sessions. When modeling the time-evolving item correlation within a session, both the user's sequential behavior (short-term) and the overall cross-session dependencies (long-term) should be taken into account~\cite{liu2018stamp,li2020time}. Therefore, it is a significant challenge to jointly integrate the intra-session item correlations and inter-session item transition patterns, into the recommendation framework in a fully adaptive manner.\\\vspace{-0.12in}

\noindent \textbf{Present Work}. Motivated by the aforementioned challenges, we propose a new multi-task learning model with \underline{M}ulti-level \underline{T}ransition \underline{D}ynamics (\model) for session-based recommendation. In our \model\ framework, we first devise a position-aware attention mechanism to jointly capture the intra-session sequential item transitions, and session-specific main purchase with the incorporation of position information. Specifically, we integrate a self-attention model with an attentive aggregation layer to capture the sequential transitional patterns of items within each individual session, without the rigid order assumption of user behavior (\ie, latent states are propagated through temporally-ordered sequences in recurrent framework). To argument the representation learning ability over individual session, an attentive summarization layer is introduced to adaptively perform pattern aggregation. In the hierarchical attentive component, we also seek to explore the item positional information under a sequential encoding module to learn the influence of time factors. Additionally, inspired by the effectiveness of mutual information maximization in prioritizing global or local structural information in feature learning~\cite{hjelm2018learning}, we model the cross-session item dependencies in a hierarchical manner, \ie, from item-level embedding learning to global graph-level representation. The developed hierarchically structured encoder via graphical mutual information maximization, endows the \model\ with the capability to incorporate inter-session transitional signals from low-level to high-level across different sessions. Source code is released at the link \emph{https://github.com/sessionRec/MTD}.

We highlight key contributions of this paper as follows:\vspace{-0.08in}

\begin{itemize}[leftmargin=*]

\item We exploit multi-level item transition dynamics in studying the session-based recommendation task. Towards this end, we propose a new recommendation framework which captures the item transition patterns, in the form of of the intra-session item dependencies, as well as the cross-session item relation structures.

\item We first develop a position-aware attentive mechanism to learn the evolving intra-session behavioral sequential signals and the summarized session-specific knowledge. Furthermore, a global context enhanced inter-session relation encoder is built upon the graph neural network paradigm, to endow \model\ for capturing the inter-session item-wise dependencies.

\item Our extensive experiments on three real-world datasets demonstrate that \model\ outperforms different types of baselines in yielding better recommendation results. Also, we show the efficiency of our developed model as compared to representative competitors and perform case studies with qualitative examples to investigate the interpretation capability of our \model\ model.
\end{itemize}

%% file: solution.tex
\section{Methodology}
\label{sec:solution}

In this section, we present the technical details of our proposed recommendation framework \model. We first formulate our studied session-based recommendation scenario as follows: Session-based recommendation aims to predict the next action of users based on their anonymous historical activity sequences (\eg, clicks or purchases). Let $S = \{v_1,...,v_m,... , v_M\}$ denote the item candidate set, where $M$ is the number of items. An anonymous session $s$ is a item sequence $s=[v_{s,1},...,v_{s,i},...,v_{s,I}]$ in a chronological order, where $v_{s,i} \in S$ denotes the $i$-th item interested by the user in the session $s$, and $I$ denotes the length of session $s$. The recommendation model outputs a list $Y=[y_1, y_2, ..., y_M]$ for each session $s$, where $y_m$ denotes the probability that the next interacted item is $v_m$. We finally make recommendations based on the top-$K$ ranked items in terms of their estimated probability values.

\subsection{Intra-Session Item Relation Learning}
To capture item transitional relationships within a session, we integrate two modules for learning the session-specific item transition patterns: (i) position-aware self-attention network for sequential transition modeling; (ii) attentive aggregation for session-specific knowledge representation.\vspace{-0.05in}

\subsubsection{\bf Self-Attentive Item Embedding Layer.} In \model\ framework, we leverage the self-attention mechanism to learn the relevance scores over historical interested items within the session and draw the sequential contextual signals. Motivated by the attentive neural network in relation learning~\cite{huang2019mist}, self-attention mechanism has been proposed to tackle various sequence modeling tasks such as machine translation~\cite{yang2019convolutional} and user behavior modeling~\cite{kang2018self}). Different from the standard attention module, self-attention could bring the benefits of capturing the relevance of past instances (\eg, words or behaviors), and refine the representation process on the single sequence at various distance~\cite{vaswani2017attention}. Following the transformer network, we build the intra-session transition modeling layer upon the dot-product attention which consists of query, key and value dimensions. The weight matrices $\textbf{W}_Q$, $\textbf{W}_K$, $\textbf{W}_V \in \mathbb{R}^{d\times d}$ respectively corresponds to the query, key, value vectors, to map initial item embeddings $\textbf{E}_s \in \mathbb{R}^{I\times d}$ of session $s$ into latent representations. The operations of self-attention network are defined as follows:
\begin{align}
\begin{bmatrix}
\textbf{Q} \\ \textbf{K} \\ \textbf{V} 
\end{bmatrix}
= \textbf{E}_s 
\begin{bmatrix}
\textbf{W}_Q \\ \textbf{W}_K \\ \textbf{W}_V 
\end{bmatrix};~~
\mathrm{Att}(\textbf{Q}, \textbf{K}, \textbf{V}) = \delta(\frac{\textbf{Q}\textbf{K}^T}{\sqrt{d}})\textbf{V}
\label{eq6}
\end{align}
\noindent where we define $\textbf{X}_s \in \mathbb{R}^{I\times d} = \mathrm{Att}(\textbf{Q}, \textbf{K}, \textbf{V})$ to represent the learned item embeddings with the modeling of pairwise relations between items $[v_{s,1},...,v_{s,i},..., v_{s,I}]$ in session $s$. $\delta(\cdot)$ denotes the softmax function and $\sqrt{d}$ is the scaling factor during the inner product operation.

We further enhance the self-attentive transition learning module with the modeling of non-linearities with the feed-forward network as shown below:
\begin{align}
\widetilde{\textbf{X}}_s = \mathrm{FFN}(\textbf{X}_s) = \varphi(\textbf{X}_s \cdot \textbf{W}_1 + \textbf{b}_1) \cdot \textbf{W}_2 + \textbf{b}_2 
\end{align}
\noindent we utilize $\varphi(\cdot)$=ReLU as the activation function. $\textbf{W}_1$, $\textbf{W}_2 \in \mathbb{R}^{d\times d}$ and $\textbf{b}_1$, $\textbf{b}_2 \in \mathbb{R}^{d}$ are trainable weight matrices and bias terms. After integrating the self-attention layer with the feed-forward network, we generate the embeddings $\widetilde{\textbf{X}}_s \in \mathbb{R}^{I\times d}$ for all items $[v_{s,1},..., v_{s,I}]$ in each session.

\subsubsection{\bf Position-aware Item-wise Aggregation Module.} We further design a position-aware attentive aggregation component to fuse the encoded item-wise relations for capturing the user main purpose within individual session $s$. We assign larger importance to the item states in which they have more contextual relations with the future interested item. In particular, for the set of items in session $s$, we learn a set of weights $\{\alpha_1$,...,$\alpha_t$,...,$\alpha_I\}$ corresponding to the set of learned item embeddings $\widetilde{\textbf{X}}_s = \{ \textbf{x}_{s,1},..., \textbf{x}_{s,i},...,\textbf{x}_{s,I} \}$. Formally, $\alpha_i$ is calculated as follows:
\begin{align}
\alpha_i = \delta (\textbf{g}^T \cdot \sigma ( \textbf{W}_3 \cdot \textbf{x}_{s,I} + \textbf{W}_4 \cdot \textbf{x}_{s,i}) )
\end{align}
\noindent where $\textbf{g} \in \mathbb{R}^{d}$ is a linear projection vector for generating the weight scalar $\alpha_i$. $\textbf{W}_3$, $\textbf{W}_4 \in \mathbb{R}^{d\times d}$. $\sigma(\cdot)$ and $\delta(\cdot)$ denotes the sigmoid and softmax function, respectively. The aggregated session representation as $\textbf{x}_s^*$, \ie, $\textbf{x}_s^* = \sum_{i=1}^I \alpha_i \cdot \textbf{x}_{s,i}$.

We further augment the intra-session item-wise fusion module with the injection of positional information, to capture the session-specific temporally-order signals of items. The dimensionality of positional representation is also set as $d$. This endows the modeling of relative positions with the incorporation of decay factor into linear transformations:
\begin{align}
\textbf{p}_s = \sum_{i=1}^I \omega_i \cdot \textbf{x}_{s,i};~~~\omega_i = \propto \mathrm{exp}(|i-I| + 1)
\end{align}
\noindent where $\textbf{p}_s$ denotes the fused representation with the preservation of relative positional information across different items. We construct a concatenated embedding for individual session of $s$ as $\textbf{q}_{s} = \textbf{W}_c [\textbf{x}_{s,I}, \textbf{x}_s^*, \textbf{p}_s]$, where $\textbf{W}_c \in \mathbb{R}^{d\times 3d}$ performs the transformation operation. After that, following the implicit feedback-based recommendation paradigm in~\cite{he2020lightgcn,wang2019neural}, we utilize the inner product between $\textbf{q}_s$ and embedding of item candidate $\textbf{v}_m$ as $\textbf{z}_{m} = \textbf{q}_s^T \textbf{v}_m$ and define our loss function of intra-session item relation learning with the cross-entropy as follows:
\begin{align}
\mathcal{L}_{in} = -\sum_n^N \textbf{y}_n \mathrm{log}(\tilde{\textbf{y}}_n)+(1 - \textbf{y}_n)\mathrm{log}(1-\tilde{\textbf{y}}_n)
\label{eq11}
\end{align}
\noindent where $\textbf{y}_n$ denotes the ground truth label of $n$-th instance and $\tilde{\textbf{y}}_n$ is the corresponding estimated result (\ie, $\tilde{\textbf{y}}_n = \delta(\textbf{z}_{n})$).

\begin{figure*}
    \centering
    \includegraphics[width=0.96\textwidth]{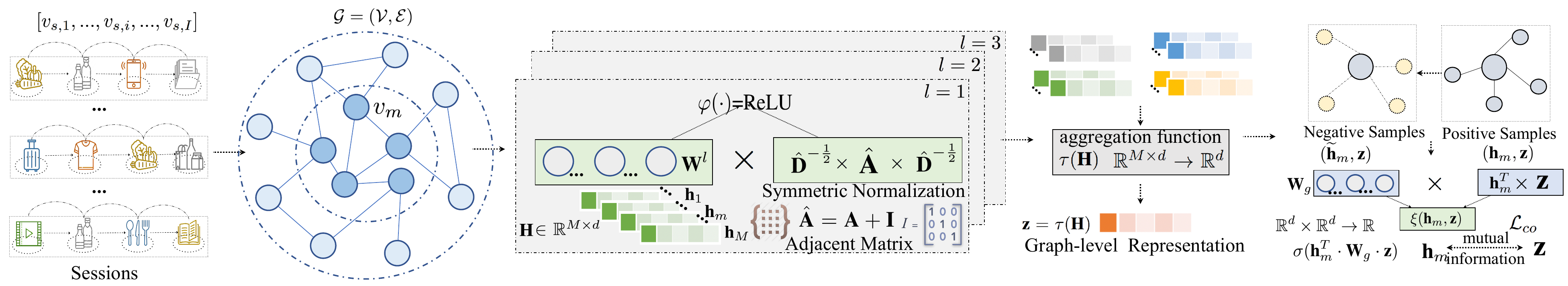}
    \vspace{-0.2in}
    \caption{Global transition dynamics modeling}
    \vspace{-0.15in}
    \label{fig:framework}
\end{figure*}

\subsection{Global Transition Dynamics Modeling}
To comprehensively capture the global cross-session transition dynamics among items, we develop a graph neural network architecture (as illustrated in Figure~\ref{fig:framework}) to inject high-order dependent signals across different sessions into session representations. In particular, we first formulate a cross-session item graph $\mathcal{G}=(\mathcal{V},\mathcal{E})$ in which nodes $\mathcal{V}$ and $\mathcal{E}$ are generated from historical sessions. Each session $s$ can be regarded as a path which starts from $v_{s,1}$ and ends at $v_{s,I}$ in graph $\mathcal{G}$. The adjacent matrix $\mathcal{A}$ is constructed where each entry $a_{m,m'}=1$ if there exists a transition relation from item $v_{m}$ to $v_{m'}$ and $a_{m,m'}=0$ otherwise.

We first propose a graph-structured message passing architecture to model the local context of transitional signals between different items. We formally define the corresponding encoding function as follows:
\begin{align}
\label{eq:patch_embed}
\textbf{H}^{(l+1)} = \varphi(\textbf{A}, \textbf{H}^l \textbf{W}^l)=\varphi(\hat{\textbf{D}}^{-\frac{1}{2}} \hat{\textbf{A}} \hat{\textbf{D}}^{-\frac{1}{2}} \textbf{H}^l \textbf{W}^l)
\end{align}
\noindent where $\textbf{H}^{(l+1)} \in \mathbb{R}^{M\times d}$ denotes the learned representations over items under the $l$-th propagation layer. With the aim of incorporating the self-propagated signals, we update the adjacent matrix with the summation of identify matrix $\textbf{I}$ and the original adjacent matrix $\textbf{A}$ as $\hat{\textbf{A}}=\textbf{A}+\textbf{I}$. Then, we further apply the symmetric normalization strategy to conduct the information aggregation as: $\hat{\textbf{D}}^{-\frac{1}{2}} \hat{\textbf{A}} \hat{\textbf{D}}^{-\frac{1}{2}}$, where $\hat{\textbf{D}}$ is the diagonal node degree matrix of $\textbf{A}$.\vspace{-0.05in}

\subsubsection{Global Dependency Representation.} After obtaining $\textbf{H}=\{\textbf{h}_{1},...,\textbf{h}_{m},...\textbf{h}_{M}\}$, we propose to capture the high-order global dependencies across correlated items from different sessions. Specifically, we first generate a fused graph-level emebdding with the aggregation function as: $\textbf{z} = \tau(\textbf{H})$ ($\mathbb{R}^{M\times d} \rightarrow \mathbb{R}^d$), where $\tau(\cdot)$ denotes the mean pooling operation. Motivated by the paradigm of global feature representation with mutual information~\cite{velivckovic2018deep,velickovic2019deep}, we enhance our cross-session item relation encoder with the global context of the mutual information between local-level embedding ($\textbf{H}$) and graph-level representation $\textbf{z}$.

We develop a classifier to perform the global dependency representation under the mutual information learning paradigm. It aims to differentiate positive ($\textbf{h}_m, \textbf{z}$) and negative instances ($\widetilde{\textbf{h}}_m, \textbf{z}$) in graph $\mathcal{G}$ by preserving the underlying cross-session item transition dynamics, the negative sample pair $(\widetilde{\textbf{h}}_m, \textbf{z})$ are generated by associating sampled item nodes with the fake embeddings based on the node shuffling strategy~\cite{velickovic2019deep}. Then, both the positive and negative instances are fed into the classifier for classification task with the encoding function $\xi(\cdot)$:
\begin{align}
\xi(\textbf{h}_{m}, \textbf{z}) = \sigma(\textbf{h}_{m}^T \cdot \textbf{W}_g \cdot \textbf{z}); \mathbb{R}^d \times \mathbb{R}^d \rightarrow \mathbb{R}
\end{align}
\noindent where $\textbf{W}_g\in \mathbb{R}^{d\times d}$ is the projection matrix. The classifier function outputs a probability score of the target node belongs to $\mathcal{G}$ given the corresponding embedding pair $(\textbf{h}_{m}, \textbf{z})$. The loss function of our graph-level global dependency representation component is defined as follows:
\begin{align}
\label{eq:dgi_loss}
\mathcal{L}_{co} &= - \frac{1}{N_{pos}+N_{neg}} \Big ( \sum_{i=1}^{N_{pos}} \rho(\textbf{h}_{m}, \textbf{z}) \cdot log \xi(\textbf{h}_{m}, \textbf{z}) \nonumber\\
&+ \sum_{i=1}^{N_{neg}} \rho(\widetilde{\textbf{h}}_{m}, \textbf{z}) \cdot log [1-\xi(\widetilde{\textbf{h}}_{m}, \textbf{z})] \Big )
\end{align}
\noindent where $\rho(\cdot)$ is an indicator function where $\lambda(\textbf{h}_{m}, \textbf{z})=1$ and $\rho
(\widetilde{\textbf{h}}_{m}, \textbf{z})=1$ corresponds to positive and negative instance pairs, respectively. We define the number of positive and negative samples as $N_{pos}$ and $N_{neg}$. By minimizing $\mathcal{L}_{co}$ (maximizing the mutual information between local-level and graph-level representations), we could generate the enhanced user representations $\textbf{H}^* \in \mathbb{R}^{M\times d}$ by encoding cross-session item transitional patterns from low-level (locally) to high-level (globally).

\subsection{Model Inference}
Based on the multi-task learning framework of \model, we define our loss function with the integration of both intra- and inter-session transition dynamics as follows:
\begin{equation}
\mathcal{L} = \mathcal{L}_{cr} + \lambda_1 \mathcal{L}_{in} + \lambda_2\Vert{{\Theta}}\Vert_2^2
\label{eq12}
\end{equation}
where ${\Theta}$ are learnable parameters. $\lambda_1$ and $ \lambda_2$ balance the losses from two module and prevent over-fitting, respectively. Since the input of cross-session relation encoder and attention network are different, we employ mini-batch Adam to optimize $\mathcal{L}_{in}$ and $\mathcal{L}_{cr}$ alternatively. We further define additional parameter $f$ to denote the training frequency of $\mathcal{L}_{in}$ optimization for loss balance. In each epoch, we first optimize the graph-structured relation encoder and initialize the item representations with the current embeddings. Note that the local representation $\textbf{H}$, which are generated by the graph neural network, implies the global transition of items. To capture the global signal in recommendation module, we update the embedding table of items with $\textbf{H}$ after the optimization step of $\mathcal{L}_{cr}$.

\noindent \textbf{Complexity Analysis of \model\ Framework.}
The intra-session item relation learning requires $O(I\times d^2+I^2\times d)$ calculations to compute the $\textbf{Q}, \textbf{K}, \textbf{V}$ and attentive embeddings $\textbf{X}_s$ in the self-attention layer. After that, the rest of the intra-session learning spends most complexity on transformations in the $d$-dimensional hidden space (\eg, the two-layer feed-forward network), which costs $O(I\times d^2)$ complexity, and results in $O(L_1\times I\times d^2+I^2\times d)$ overall complexity. Here, $L_1$ denotes the number of $d\times d$ transformations. Furthermore, the graph-based inter-session item transition modeling component requires $O(|\textbf{A}|\times d + M\times d^2)$ complexity for message passing and embedding transformation, where $|\textbf{A}|$ denotes the number of neighboring item pairs.


%% file: eval.tex
\section{Evaluation}
\label{sec:eval}

In this section, we perform extensive experiments on three publicly available real-life recommendation datasets and compare \emph{\model} with various state-of-the-art techniques. Particularly, we aim to answer the following research questions:

\begin{itemize}[leftmargin=*]
\item \textbf{RQ1}: Does \emph{\model} consistently outperform other baselines by yeilding better recommendation results?
\item \textbf{RQ2}: How do different sub-modules in our \emph{\model} framework affect the recommendation performance?
\item \textbf{RQ3}: What is the influence of hyperparameter settings in \emph{\model} for the model performance?
\item \textbf{RQ4}: How is the model interpretation capability of \emph{\model}?
\item \textbf{RQ5}: How is the computational cost of \emph{\model} method ?
\end{itemize}

\begin{table}
\centering
{
\footnotesize
\begin{tabular}{| c | c | c | c |}
\hline
Dataset & Yoochoose & Diginetica & RetailRocket\\
\hline
\# Train Sessions & 369,859 & 719,470 & 433,648 \\
\hline
\# Test Sessions  & 55,400  & 60,858  & 15,132 \\
\hline
\# All Items      & 17,376  & 43,097  & 36,968 \\
\hline
Average Length    & 6.15    & 5.13    & 9.93   \\
\hline
\end{tabular}
}
\vspace{-0.1in}
\caption{Statistics of the experimented datasets.}
\vspace{-0.2in}
\label{tab:data}
\end{table}

\subsection{Experimental Settings}
\subsubsection{\bf Data Description.}
The data statistics with training/test detailed split settings are shown in Table~\ref{tab:data}. We present the details of experimented datasets as below:\\\vspace{-0.1in}

\noindent \textbf{Yoochoose Data}\footnote{http://cikm2016.cs.iupui.edu/cikm-cup}. This data comes from an online retailing site to log half year of user clicks (released by Recsys'15 Challenge). Following the pre-processing strategies in~\cite{li2017neural,liu2018stamp}, the sessions with the length of $\geq 2$ and items with the appearing frequency of $\geq 5$ are kept in the training and test set.\\\vspace{-0.1in}

\noindent \textbf{Diginetica Data}\footnote{http://2015.recsyschallenge.com/challenge.html}. This data is collected from the CIKM Cup platform 2016 which records the user clicks from the time period of six months. To be consistent with the settings in~\cite{wu2019session,liu2018stamp}, we do not include the sessions that contains single clicked item. Sessions in the test set are generated from the last week. \\\vspace{-0.1in}

\noindent \textbf{Retailrocket Data}\footnote{https://www.kaggle.com/retailrocket/ecommerce-dataset}. It contains the user browse data from another e-commerce company. Following the same settings in~\cite{xu2019graph}, we filter out the items with the browsed frequency less than 5 and sessions with the length of less than 2. We set the data from the last week for test and the remaining part for training. 

\subsubsection{\bf Evaluation Metrics.}
We leverage two metrics which are widely adopted in the session-based recommendation applications: \textbf{Precision@$K$} (Pre@$K$) and \textbf{Mean Reciprocal Rank@$K$} (MRR@$K$). Following the same rubric in~\cite{wu2019session,li2017neural}, MRR@$K$=0 if the first correctly recommended items is not among the top-$K$ ranked items. Note that larger Pre@$K$ and MRR@$K$ scores indicate better recommendation performance.

\subsubsection{\bf Compared Methods.} In our experiments, we consider the following baselines for performance comparison.
\begin{itemize}[leftmargin=*]
\item \textbf{POP}: it explores users' past interested items and makes recommendations with the identified most frequent items.
\item \textbf{S-POP}: it recommends the most popular items to users by considering their activities from the current session.

\item \textbf{ItemKNN}~\cite{davidson2010youtube}: it considers the item correlations using $k$-nearest neighbors algorithm based on items' cosine similarity.

\item \textbf{GRURec}~\cite{hidasi2015session}: it is a representative session-based recommendation approach using the gated recurrent unit to encode the transitional regularities.

\item \textbf{NARM}~\cite{li2017neural}: it is a neural attention model to argument recurrent network for session representations, by attending deferentially to sequential items.
\item \textbf{STAMP}~\cite{liu2018stamp}: this approach is an attention model to capture user's temporal interests from historical clicks in a session.
\item \textbf{SASRec}~\cite{kang2018self}: this method is built upon the self-attention architecture to model the long-term item transition dynamics.

\item \textbf{SR-GNN}~\cite{wu2019session}: it proposes a graph neural network model to encode item transitions within a session to generate item embedding.

\item \textbf{CSRM}~\cite{wang2019collaborative}: it integrates the inner memory encoder through an outer memory network by considering correlations between neighborhood sessions.
\item \textbf{CoSAN}~\cite{luocollaborative}: it designs self-attention networks to model the collaborative feature information of items from neighborhood sessions.
\end{itemize}

\subsection{Parameter Settings}
Our implement is based on Tensorflow. The embedding dimensionality $d$ is set as 100. We assign the regularization penalty $\lambda_2 = 10^{-6}$. All models are optimized using the Adam optimizer with the batch size and learning rate as 512 and $1e^{-3}$, respectively. The training frequency $f$ in each epoch is set as 1, 4, 6 corresponding to the Yoochoose, Diginetica, Retailrocket, respectively. Furthermore, the dropout technique is applied in the training phase to alleviate the overfitting issue, with the ratio of 0.2. Experiments of most baselines are conducted with their release source code.


\begin{table*}
\centering
\footnotesize
 \begin{tabular}{l|c|c|c|c|c|c|c|c|c|c|c|c}
\hline
Data & Metric & ~POP~ & S-POP & It-KNN & GRURec & NARM & STAMP & SASRec & SR-GNN & ~CSRM & CoSAN & \emph{\model} \\
\hline
\multirow{2}{*}{Digi} 
                            & Pre & 0.58 & 20.66 & 26.46 & 20.31 & 36.72 & 37.05 & 38.42 & 38.40 & 38.56 & 37.58 & \textbf{40.22}\\
                            & MRR & 0.19 & 13.59 & 10.91 & 7.78  & 15.00 & 16.05 & 16.27 & 17.04 & 16.23 & 15.57 & \textbf{17.58}\\
\hline
\multirow{2}{*}{Yooc}  
                            & Pre & 4.59 & 28.61 & 43.40 & 55.13 & 60.19 & 58.79 & 60.42 & 60.84 & 60.46 & 61.01 & \textbf{61.83}\\
                            & MRR & 1.51 & 18.45 & 21.39 & 25.76 & 29.03 & 29.44 & 30.47 & 30.57 & 30.37 & 30.21 & \textbf{30.83}\\
\hline
\multirow{2}{*}{Reta} 
                              & Pre & 1.59 & 29.67 & 21.41 & 31.01 & 44.74 & 43.14 & 46.39 & 44.88 & 47.21 & 45.83 & \textbf{47.93}\\
                              & MRR & 0.44 & 21.51 & 9.78  & 15.37 & 25.54 & 26.65 & 26.74 & 26.95 & 27.14 & 26.01 & \textbf{28.51}\\
\hline
\end{tabular}
\vspace{-0.1in}
\caption{Recommendation performance comparison of all methods in terms of Pre@$10$ and MRR@$10$.}
\label{tab:result_across}
\vspace{-0.1in}
\end{table*}

\begin{table}
\centering
\footnotesize
 \vspace{-0.05in}
 \begin{tabular}{l|c|c|c|c|c}
\hline
Data & Metric & SR-GNN & ~CSRM & CoSAN & \emph{\model} \\
\hline
\multirow{3}{*}{Digi} & Pre@5  & 27.15 & 26.38 & 25.72 & \textbf{28.29}\\
                            & Pre@10 & 38.40 & 38.56 & 37.58 & \textbf{40.22}\\
                            & Pre@20 & 51.57 & 52.56 & 50.94 & \textbf{53.92}\\
\hline
\multirow{3}{*}{Reta} & Pre@5 & 37.38 & 38.65 & 37.07 & \textbf{39.64}\\
                              & Pre@10 & 44.88 & 47.21 & 45.83 & \textbf{47.93}\\
                              & Pre@20 & 52.27 & 55.04 & 54.87 & \textbf{55.95}\\
\hline
\end{tabular}
\vspace{-0.1in}
\caption{Evaluation results with different top-$K$ values.}
\label{tab:result_vary_k}
\vspace{-0.1in}
\end{table}

\subsection{Performance Validation (RQ1)}

We present evaluation results of all methods on different datasets in Table~\ref{tab:result_across}, and show the performance of several recent baselines when varying the value of top-$K$ in Table~\ref{tab:result_vary_k}. We can observe that \emph{\model} consistently outperforms other baselines in most cases on different datasets, which justifies the effectiveness of our model in comprehensively capturing multi-level transition dynamics from intra-session and inter-session relations in a hierarchical manner.

The naive frequency (POP and S-POP) and similarity (ItemKNN) based recommendation approaches perform much worse than other baselines due to their limitations in capturing the dynamic sequential patterns of item transitions. Additionally, the attention-based recommendation techniques (NARM and STAMP) outperform the mere RNN approach (GRU4REC)--considering singular level of item sequential relations. However, the significant improvement between \emph{\model} and attentive recommendation model suggests that only considering the intra-session item transitions is insufficient to fully capture the complex item transition dynamics from both local and global perspectives. While SR-GNN tries to encode the long-term item dependencies using the graph neural network, it yields suboptimal results because its failure in learning cross-session dependency.

\subsection{Model Ablation and Effect Analyses (RQ2)}
We consider several model variants to investigate the efficacy of key modules in our learning framework of \emph{\model}.\\\vspace{-0.1in}

\noindent \textbf{Effect of Hierarchical Attention Network}.
We design two contrast models: i) \emph{\model}-va generates the session-level embeddings with the vanilla attention layer; ii) \emph{\model}-at further incorporates the temporal factor into the \emph{\model}-va method.\\\vspace{-0.1in}

\noindent \textbf{Effect of Cross-Session Dependency Encoder}. i) \emph{\model}-lo only encodes the local-level item transition patterns without the cross-session dependency encoder; ii) \emph{\model}-ga replaces our graph-structured hierarchical relation encoder with the graph attention network operated on all relevant sessions.

\begin{figure}[h]
	\centering
	\vspace{-0.1in}
	\subfigure[][Diginetica]{
		\centering
		\includegraphics[width=0.13\textwidth]{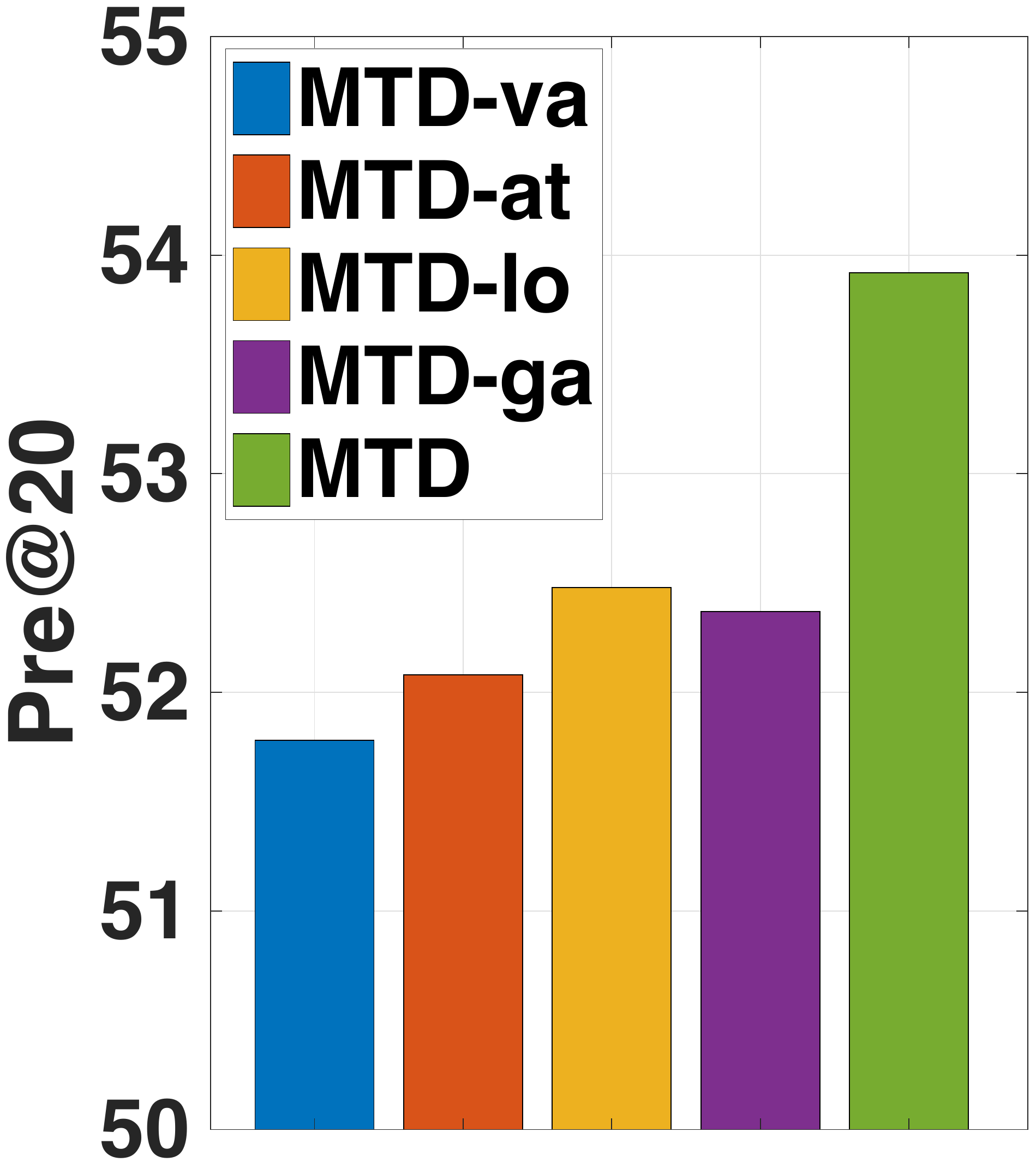}
		\label{fig:beh_buy_hr}
	}
	\subfigure[][Yoochoose]{
		\centering
		\includegraphics[width=0.14\textwidth]{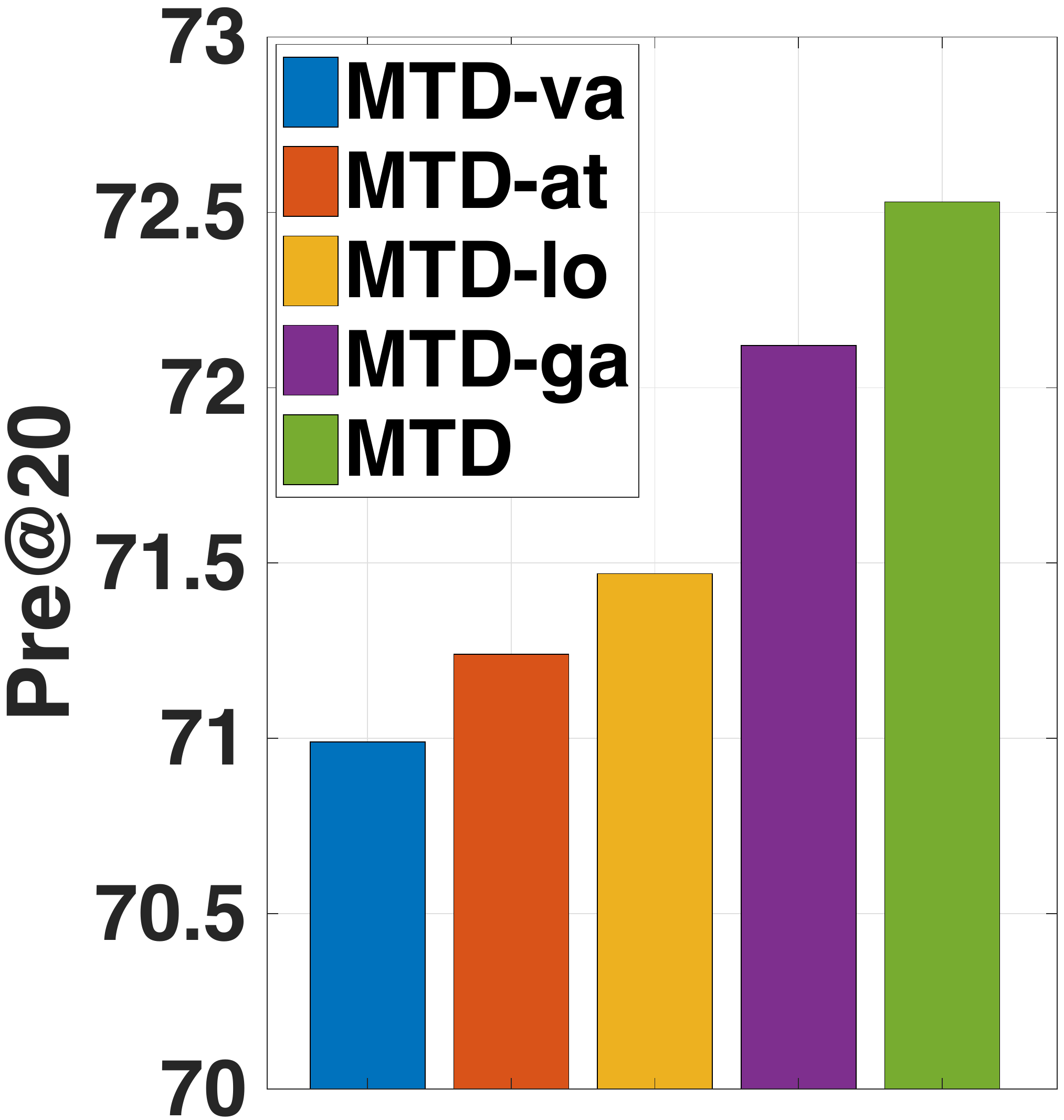}
		\label{fig:beh_click_hr}
	}
	\subfigure[][Retailrocket]{
		\centering
		\includegraphics[width=0.14\textwidth]{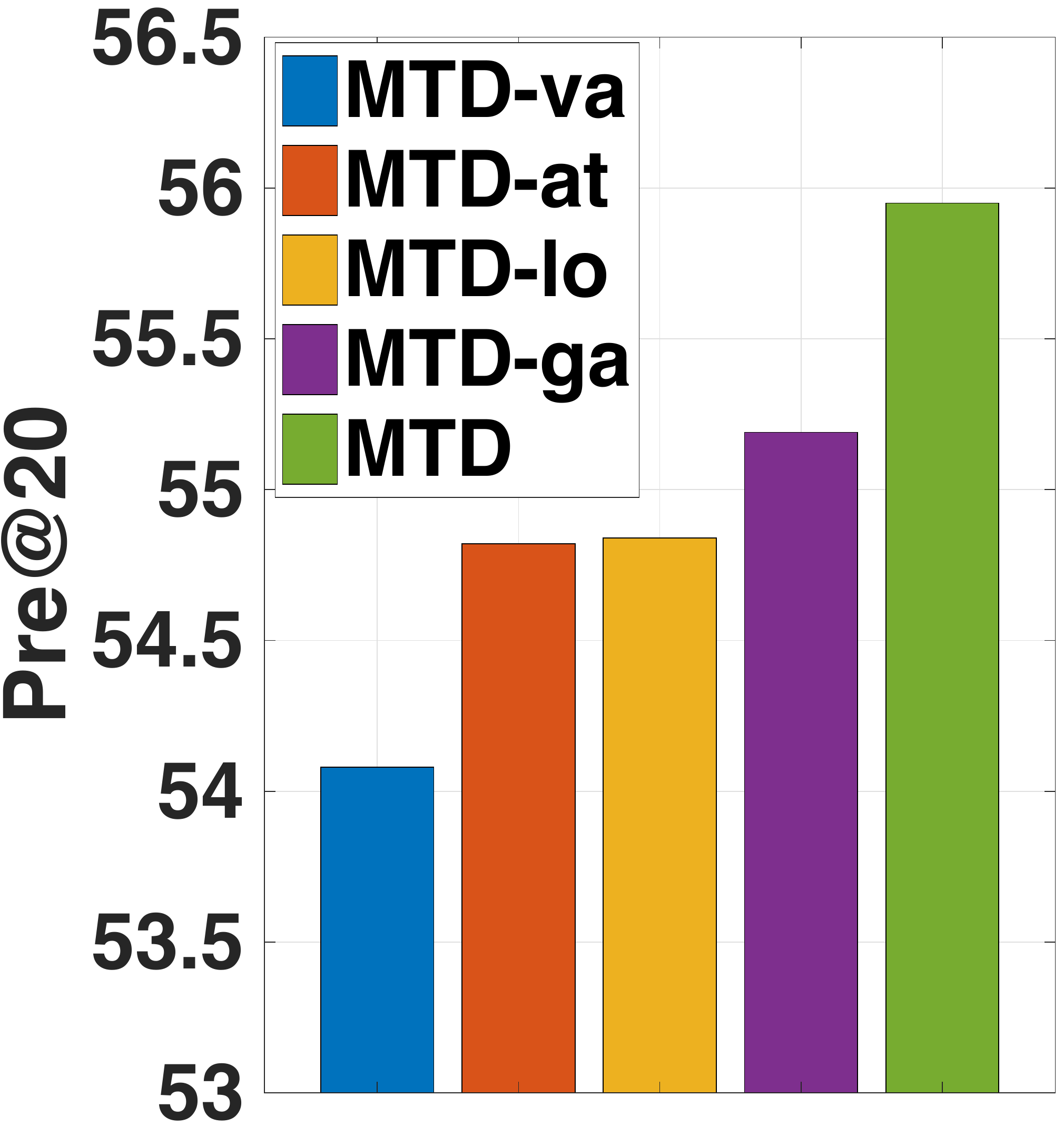}
		\label{fig:beh_click_ndcg}
	}
	\vspace{-0.1in}
	\caption{Model ablation study of \emph{\model}.}
	\label{fig:model_ablation}
	\vspace{-0.05in}
\end{figure}

We report the results in Figure~\ref{fig:model_ablation} and observe that \emph{\model} outperforms all other variants on all datasets in terms of $Pre$@$K$ and $MRR$@$K$ under $K=20$, which justifies the effectiveness of the design of individual component in our \emph{\model} framework. In particular: (1) The performance gap among \emph{\model}-va, \emph{\model}-at, and \emph{\model}-lo shows the effectiveness of our position-aware hierarchical attention network in modeling the local item transitions. (2) Without the consideration of cross-session item dependencies, \emph{\model}-lo performs worse than \emph{\model}. It suggests the necessity of modeling the inter-session item correlations based on our developed graph-structured framework; (3) While the graph attention network (\emph{\model}-ga) could learn global-level item relations, it still falls behind \emph{\model} since it does not capture the hierarchical informativeness across relevant sessions.

\subsection{Hyperparameter Study of MTD (RQ3)}
\begin{figure}
\vspace{-0.1in}
    \centering
    \begin{adjustbox}{max width=1.0\linewidth}
    \input{./fig/parameter}
    \end{adjustbox}
    \vspace{-0.20in}
    \caption{Hyper-parameter study of \emph{\model}.}
    \vspace{-0.15in}
    \label{fig:hyperparam_study}
\end{figure}
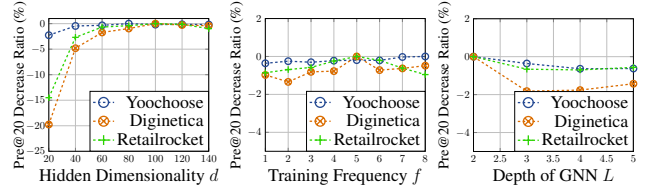

We further investigate the hyperparameter sensitivity of our \emph{\model} (as shown in Figure~\ref{fig:hyperparam_study}) and summarize the following observations. To save space and integrate results on different datasets with different performance scales into the one figure, we set y-axis as the performance degradation ratio compared to the best performance. \\\vspace{-0.1in}

\noindent (1) \textbf{Effect of Hidden Dimensionality $d$}. The performance saturates as the hidden dimensionality $d$ reaches around 100. This is because a larger dimensionality $d$ brings a stronger representation ability at the early stage, but might lead to overfitting as the continuously increasing of $d$.\\\vspace{-0.1in}

\noindent (2) \textbf{Impact of Training Frequency $f$}. We perform the training frequency study by varying $f$ from 1 to 8, and could notice that a large value of $f$ ($\geq 5$) will degrade the performance by misleading the objective function optimization.\\\vspace{-0.1in}

\noindent (3) \textbf{Influence of Depth in Graph Neural Architecture}.
Stacking more graph convolution layers with the adjacent matrix-based aggregation will involve more redundant information of high-order connectivity, which hinders the learning process of global item relational structures in \emph{\model}. This observation also suggests the rationality of our designed graph neural component in simplifying and powering the cross-session item dependency learning, via the exploration of mutual relations between low-level item embeddings and high-level graph representation.

\subsection{Case Studies: Model Interpretation (RQ4)}
\begin{figure}[h!]
    \vspace{-0.10in}
    \includegraphics[width=0.47\textwidth]{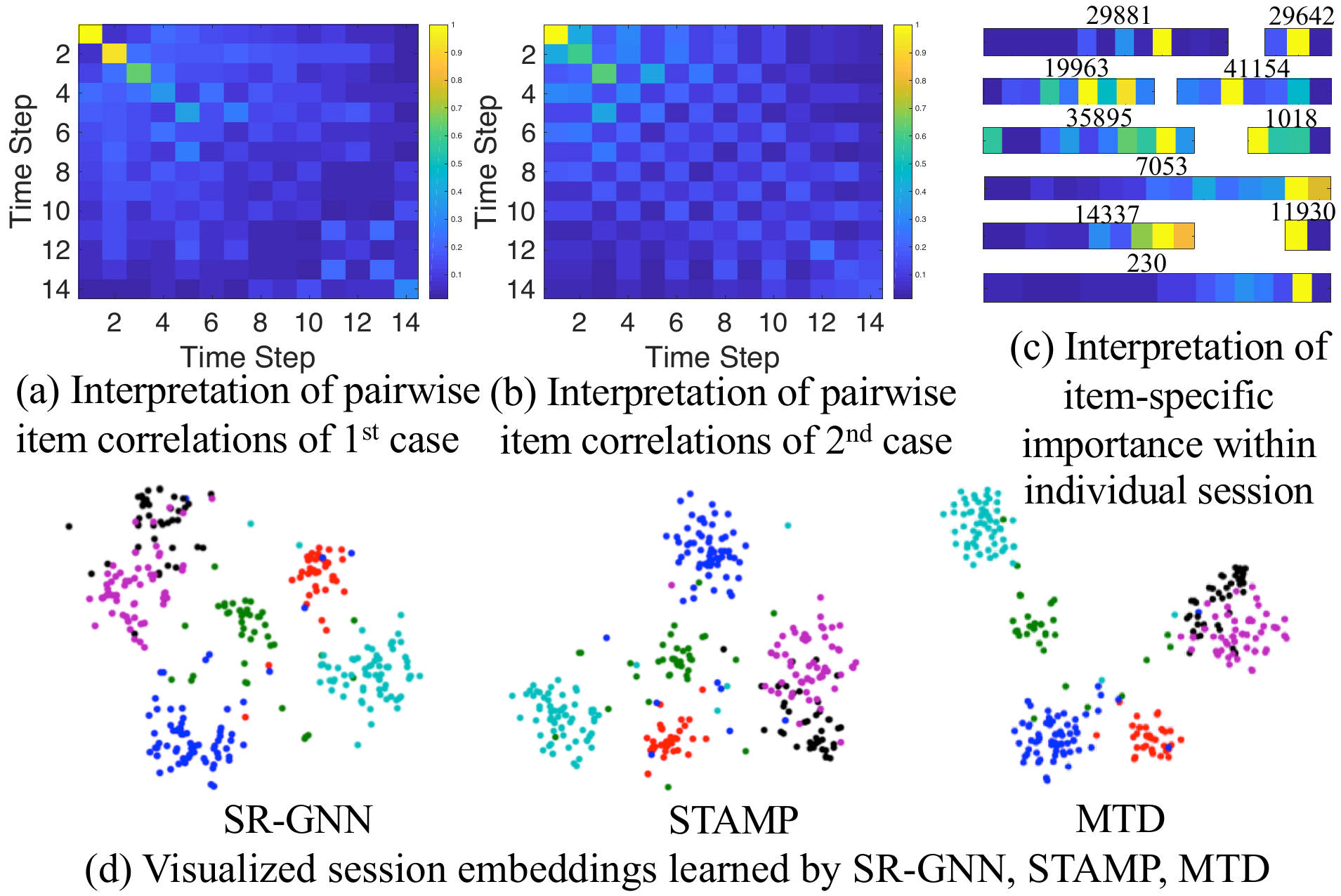}
    \vspace{-0.10in}
    \caption{Case study of \emph{\model} framework}
    \label{fig:case_study}
    \vspace{-0.15in}
\end{figure}

\noindent {\bf Hierarchical Relation Interpretation across Items.}
We visualize the hierarchical item relations with quantitative weights learned from our intra-session attention network on Diginetica. Figure~\ref{fig:case_study} (a) and Figure~\ref{fig:case_study} (b) show the encoded pairwise item correlations in modeling the intra-session sequential patterns of two sampled sessions across different time steps. From Figure~\ref{fig:case_study} (c), we can observe that different items contribute differently to summarize the session-specific main purchase with hidden representations.\\\vspace{-0.1in}

\noindent{\bf Visualizations of Learned Session Embeddings.}
We further visualize the projected session representations by our \emph{\model} and two state-of-the-arts: SR-GNN and STAMP (as shown in Figure~\ref{fig:case_study} (d)). We randomly sample 180 session instances and label each one with its corresponding next clicked item (ground truth). It is easy to see that embeddings of sessions with the same label (6 classes and each one is represented with the same color) cluster closely and can be better distinguished by \emph{\model} as compared to other two methods. This demontrates the effectiveness of our learned item transitional patterns with session embeddings.

\subsection{Model Scalability Study (RQ5)}
\label{sec:efficiency}
Since efficiency is a key factor in many real-life recommendation applications, we finally investigate the computational cost (measured by running time of individual epoch) of our \emph{\model} and other state-of-the-art recommendation models. Our experiments are conducted on different datasets are summarized in Table~\ref{tab:time}. From the evaluation results, we can observe that \emph{\model} outperforms most competitive baselines with different deep neural network architectures (\eg, attention mechanisms and graph-based message passing frameworks). Particularly, SR-GNN involves much computation cost in the gating mechanisms from neural network over each constructed session graph. Additionally, it is time-consuming to discover collaborative neighborhood sessions for each batch during the training phase of CSRM method. In the occasional cases that \emph{\model} miss the best performance (as compared to a streaming algorithm STAMP--only using attention mechanism for transition aggregation), \emph{\model} still achieves competitive model efficiency. Overall, the proposed \emph{\model} is efficient and scalable for large-scale session-based recommendation applications.

\begin{table}[h]
\vspace{-0.05in}
    \centering
    \footnotesize
    \vspace{-0.1in}
    \begin{tabular}{lccc}
         \toprule
         Models & Yoochoose & Diginetica & RetailRocket \\
         \hline
         NARM   & 35 & 66 & 81\\
         STAMP  & 9  & 24 & 14\\
         SASRec & 18 & 28 & 42\\
         TiSA   & 82 & 160 & 100\\
         SR-GNN & 1401 & 2586 & 2502\\
         CSRM   & 530  & 556  & 228\\
         \hline
         \emph{\model} & 24 & 40 & 53 \\
         \hline
    \end{tabular}
    \caption{Computational time cost (seconds) investigation.}
    \vspace{-0.1in}
    \label{tab:time}
\end{table}

%% file: fig/parameter.tex
\begin{tikzpicture}
\begin{axis}[
    xlabel={\Large{Hidden Dimensionality $d$}},
    ylabel={Pre@20 Decrease Ratio (\%)},
    xmin=20,xmax=140,
    ymin=-25,ymax=1,
    legend columns=1,
    legend cell align=right,
    grid=both,
    every axis plot/.append style={ultra thick},
    every tick label/.append style={scale=1.3},
    label style={scale=1.8},
    legend style={
        nodes={scale=1.5, transform shape},
        legend image post style={scale=1.5},
        },
    legend style={at={(1,0)},anchor=south east},
    every axis plot post/.append style={
        every mark/.append style={scale=2}
    }
]
\addplot[color={rgb:blue,4;green,2;red,1}, mark=o, dashed, mark options={solid}]
table[x=para, y=Yoochoose_pre] {dimensionality.txt};
\addplot[color={rgb:red,4;green,1;yellow,2}, mark=otimes, dashed, mark options={solid}]
table[x=para, y=Diginetica_pre] {dimensionality.txt};
\addplot[color={rgb:green,5;red,1}, mark=+, dashed, mark options={solid}]
table[x=para, y=Retailrocket_pre] {dimensionality.txt};
\legend{\LARGE Yoochoose, \LARGE Diginetica, \LARGE Retailrocket};
\end{axis}
\end{tikzpicture}


\begin{tikzpicture}
\begin{axis}[
    xlabel={\Large{Training Frequency $f$}},
    ylabel={Pre@20 Decrease Ratio (\%)},
    xmin=1,xmax=8,
    ymin=-5,ymax=2,
    legend columns=1,
    legend cell align=right,
    grid=both,
    every axis plot/.append style={ultra thick},
    every tick label/.append style={scale=1.3},
    label style={scale=1.8},
    legend style={
        nodes={scale=1.5, transform shape},
        legend image post style={scale=1.5},
        },
    legend style={at={(1,0)},anchor=south east},
    every axis plot post/.append style={
        every mark/.append style={scale=2}
    }
]
\addplot[color={rgb:blue,4;green,2;red,1}, mark=o, dashed, mark options={solid}]
table[x=para, y=Yoochoose_pre] {frequency.txt};
\addplot[color={rgb:red,4;green,1;yellow,2}, mark=otimes, dashed, mark options={solid}]
table[x=para, y=Diginetica_pre] {frequency.txt};
\addplot[color={rgb:green,5;red,1}, mark=+, dashed, mark options={solid}]
table[x=para, y=Retailrocket_pre] {frequency.txt};
\legend{\LARGE Yoochoose, \LARGE Diginetica, \LARGE Retailrocket};
\end{axis}
\end{tikzpicture}


\begin{tikzpicture}
\begin{axis}[
    xlabel={\Large{Depth of GNN $L$}},
    ylabel={Pre@20 Decrease Ratio (\%)},
    xmin=2,xmax=5,
    ymin=-5,ymax=2,
    legend columns=1,
    legend cell align=right,
    grid=both,
    every axis plot/.append style={ultra thick},
    every tick label/.append style={scale=1.3},
    label style={scale=1.8},
    legend style={
        nodes={scale=1.5, transform shape},
        legend image post style={scale=1.5},
        },
    legend style={at={(1,0)},anchor=south east},
    every axis plot post/.append style={
        every mark/.append style={scale=2}
    }
]
\addplot[color={rgb:blue,4;green,2;red,1}, mark=o, dashed, mark options={solid}]
table[x=para, y=Yoochoose_pre] {depth_GNN.txt};
\addplot[color={rgb:red,4;green,1;yellow,2}, mark=otimes, dashed, mark options={solid}]
table[x=para, y=Diginetica_pre] {depth_GNN.txt};
\addplot[color={rgb:green,5;red,1}, mark=+, dashed, mark options={solid}]
table[x=para, y=Retailrocket_pre] {depth_GNN.txt};
\legend{\LARGE Yoochoose, \LARGE Diginetica, \LARGE Retailrocket};
\end{axis}
\end{tikzpicture}


%% file: relate.tex
\section{Related Work}
\label{sec:relate}

\noindent \textbf{Session-based Recommender Systems}. To model sequential patterns of user behaviors, many recommender systems have been proposed to predict future interactions based on users' historical observations~\cite{huang2019online}. In recent years, many session-based recommendation techniques have been developed based on various neural network architectures~\cite{qiu2020exploiting}. Particularly, one intuitive approach is to apply the recurrent neural network (\eg, GRU) for modeling the item sequential correlations~\cite{hidasi2015session}. Furthermore, attention mechanisms have been adopted for pattern aggregation through relation weight learning, such as NARM~\cite{li2017neural} and STAMP~\cite{liu2018stamp}. Different from the method~\cite{xujoint} which replies on the random walk-based skipgram model for capturing the dependency, we leverage graph neural networks to consider the global item dependency across different sessions. Another paradigm of session-based recommendation models lie in utilizing graph neural networks to capture the graph-structured item dependencies, such as attributed graph neural network for streaming recommendation~\cite{qiu2020gag} and graph-based message passing architectures~\cite{wu2019session}. Different from the above work, our \model\ framework aims to jointly captures the local and global item transitional signals in a hierarchical manner. \\\vspace{-0.1in}

\noindent \textbf{Graph Neural Networks for Recommendation.} Recently emerged graph neural networks shine a light on performing information propagation over user-item graph for recommendation. Inspired by the graph convolution, several efforts have been devoted to capturing collaborative signals from the graph-based interacted neighbors, such as and LightGCN~\cite{he2020lightgcn} and PinSage~\cite{ying2018graph}. Additionally, graph neural networks have also been integrated for recommendation to aggregate external knowledge from user side~\cite{aaaisocial} or item side~\cite{wang2019explainable}. In this work, we propose to capture cross-session item dependencies in a hierarchical manner upon a global context enhanced graph network.

%% file: conclusion.tex
\section{Conclusion}
\label{sec:conclusion}

This work develops a new graph learning framework--\model, which aims to inject multi-level transition dynamics into the session-based recommendation. By integrating a position-aware dual-stage attention network and graph hierarchical relation encoder, \model\ not only models the intra-session sequential transitions, but also derives the high-order item relationships across sessions. Experimental results on different real-world datasets show that \model\ is superior to many state-of-the-art baselines. In the future, we plan to incorporate item content information (\eg, item text description or reviews) into \model\ to deal with external attributes in learning semantic-aware item transitions.